\documentclass[preprint,floats,aps,12pt]{revtex4}
\usepackage{times} 
\usepackage{amssymb}
\usepackage{amsmath}
\ifx\pdfoutput\undefined
\usepackage{graphicx}     
\else
\usepackage{epstopdf}
\usepackage[pdftex]{graphicx}
\usepackage[pdftex]{hyperref}
\hypersetup{a4paper,
    pdftitle={Thermodynamically consistent description of the hydrodynamics of free surfaces covered by
  insoluble surfactants of high concentratio}
    pdfauthor={Uwe Thiele, Andrew J. Archer and Mathis Plapp},
pdfproducer={lateX},
pdfview=FitV,       
pdfstartview=FitB,
linkcolor=blue,     
pagecolor=blue,     
urlcolor=red,      
breaklinks=true,    
colorlinks=true,
citebordercolor=0 0 0,  
citecolor=blue,  
filebordercolor=0 0 0,
linkbordercolor=0 0 0,
menubordercolor=0 0 0,
pagebordercolor=0 0 0,
urlbordercolor=0 0 0,
pdfhighlight=/I,
pdfborder=0 0 0,   
backref=false,
pagebackref=false,
bookmarks=true,
bookmarksopen=true,
bookmarksnumbered=true
}
\fi
\ifx\pdfoutput\undefined
\DeclareGraphicsExtensions{.eps}
\else
\DeclareGraphicsExtensions{.jpg, .pdf, .tif}
\fi
\usepackage{color}
\newcommand{\mylab}[1]{\label{#1}}
\usepackage{ulem}
\newcounter{countuwe}

\renewcommand{\vec}[1]{\mathbf{#1}}
\newcommand{\vecg}[1]{\boldsymbol{#1}}
\newcommand{\tens}[1]{\mathbf{\underline{#1}}}

\newcommand{\rr}{\mathbf{r}}
\begin{document}
\title{Thermodynamically consistent description of the hydrodynamics of free surfaces covered by
  insoluble surfactants of high concentration}
\author{Uwe Thiele}
\email{u.thiele@lboro.ac.uk}
\homepage{http://www.uwethiele.de}
\author{Andrew J. Archer}
\email{a.j.archer@lboro.ac.uk}
\affiliation{Department of Mathematical Sciences, Loughborough University,
Loughborough, Leicestershire, LE11 3TU, UK}
\author{Mathis Plapp}
\affiliation{Physique de la Matiere Condensee, Ecole Polytechnique, CNRS, 91128 Palaiseau, France}
\begin{abstract}
  In this paper we propose several models
  that describe the dynamics of liquid films which are covered by a high
  concentration layer of insoluble surfactant.  First, we briefly review
  the `classical' hydrodynamic form of the coupled evolution equations
  for the film height and surfactant concentration that are well
  established for small concentrations. Then we re-formulate the
  basic model as a gradient dynamics based on an underlying free
  energy functional that accounts for wettability and
  capillarity. Based on this re-formulation in the framework of
  nonequilibrium thermodynamics, we propose extensions of the basic
  hydrodynamic model that account for (i) nonlinear equations of
  state, (ii) surfactant-dependent wettability, (iii) surfactant phase
  transitions, and (iv) substrate-mediated condensation. In passing, we
  discuss important differences to most of the models found in the literature.
\end{abstract}
%
%
\maketitle
%
%
\section{Introduction} 
\mylab{sec:intro}

Small volumes of simple and complex fluids that occur naturally in
biological contexts or that are employed in modern technology, such as
e.g., in microfluidics, are often (partly) confined by a free surface
that may be covered by surface active agents. These so-called
surfactants may be tensids, lipids, certain nano-particles, or
particular polymeric compounds. Because they decrease the surface tension of the 
free surface, gradients in their concentration correspond to
gradients in the surface tension. These gradients result in tangential forces at the
free surface that drive flows in the bulk liquid. This corresponds
to the so-called solutal Marangoni effect, that is e.g., responsible
for the tears of wine \cite{Thom1855,ScSt60}.

All surface active agents are to some extent soluble in the bulk
liquid, implying that a complete dynamical model needs to describe the
motion of the bulk liquid, bulk concentration of surfactant, the
surface concentration of surfactant and the adsorption/desorption
processes that exchange surfactant molecules between the bulk liquid and the
free surface. However, for many practically important surfactants, the
bulk solubility is actually very small. Then one speaks of ``insoluble
surfactants'' and only considers the dynamics of the surfactant that
is adsorbed at the free surface. Here, we restrict our
attention to such insoluble surfactants at concentrations
at which no micelles are formed in the bulk liquid \cite{Atde10}.

The governing transport equations that relate the material properties
of the insoluble surfactant and the resulting hydrodynamic flow are
well established for low values of the surfactant surface coverage $\Gamma$
\cite{ODB97,CrMa09,Leal07}.
 In this case, the linear equation of state
\begin{equation}
\gamma(\Gamma)=\gamma_0+\gamma_\Gamma \Gamma
\mylab{eq:state-lin}
\end{equation}
describes how the surface tension deviates from its reference value
 $\gamma(0)\equiv\gamma_0$, for a bare free surface. The coefficient
$\gamma_\Gamma$ is a material constant that is negative for most
combinations of liquid and surfactant. The resulting tangential
Marangoni force at the free surface is $\nabla_s\gamma=\gamma_\Gamma
\nabla_s\Gamma$ where $\nabla_s=(\tens{I}-\vec{n}\vec{n})\cdot\nabla$
is the derivative along the free surface and $\vec{n}$ is the unit normal
vector.  For any (linear or nonlinear) equation of state, the
surface tension gradient $\nabla_s\gamma$ enters the tangential stress
boundary condition of the momentum transport equation. The latter is
accompanied by a transport equation for $\Gamma$ that
accounts for advective and diffusive transport of the surfactant
\cite{Ston90,WRM96}.  The resulting system of equations may be
simplified in order to apply them to various physical situations such as, for example, the
dynamics of surfactant-laden drops or bubbles immersed in
(another) liquid \cite{KvM04,Hame08},
free-standing soap films \cite{DGC94}, liquid bridges covered by a
surfactant monolayer \cite{LFB06},
surfactant-covered vertical falling liquid films \cite{JiSe94}, films
on horizontal solid substrates \cite{JeGr92,ODB97,CrMa09}, and drawn
meniscii \cite{Sche10}. In
particular, the latter geometry allows for an asymptotic treatment
which results in a long-wave or lubrication description of the dynamics, via
two coupled evolution equations for the film height and the surfactant
surface coverage \cite{ODB97,CrMa09}.  In the following, we focus on
this geometry, but we should emphasise that our main arguments also apply to the
general case.

Many works only treat the case of low surfactant surface coverages and
employ the linear equation of state in Eq.~\eqref{eq:state-lin}. The surface
tension driven flow is then said to result from a linear solutal
Marangoni effect. However, there is a growing literature where a
similar approach is used to treat the dynamics of free surfaces covered by
large concentrations of insoluble surfactants. It is common
practice to replace the linear equation of state~\eqref{eq:state-lin}
by a nonlinear one and leave all other terms in the dynamical equations
unchanged. We
argue below that this may result in governing equations that are
thermodynamically inconsistent, since one must also amend the
surfactant surface diffusion term.
There should also be amendments to the basic equations which describe the influence of the
surfactant coverage on wettability close to three phase contact lines of
very thin films, the effects of phase transitions
in the surfactant layer at high concentrations, 
and also the influence of a nearby solid substrate
on such phase transitions that may result in substrate mediated condensation
(surfactant aggregation). The approach that we propose in this paper allows one to deal with
all these cases in a thermodynamically consistent manner.

The structure of the paper is as follows: First, we review in section~\ref{sec:filmsurf-class} the
`classical' thin-film hydrodynamic coupled equations of motion for a
thin liquid film covered by a low concentration surfactant. Then, in a
preparatory step we decouple and re-formulate
the two individual equations in a ``thermodynamic form''. In particular,
section~\ref{sec:pure} gives the gradient formulation of the evolution
equation for a thin-film of pure liquid on a solid substrate, for the
case where capillarity and wettability are the dominant influences,
while section~\ref{sec:diff} briefly reviews the classical diffusion
equation and places it in the thermodynamic context that we
employ. In section~\ref{sec:filmsurf-graddyn} the full
coupled system is re-formulated as a gradient dynamics based on an
underlying free energy functional.  This thermodynamic form is used in
section~\ref{sec:filmsurf-ext} to extend the thin-film model to
consistently account for (i) nonlinear equations of state, (ii)
surfactant-dependent wettability, (iii) surfactant phase transitions,
and (iv) substrate mediated condensation. We also note some
differences to the models in the literature. Finally,
section~\ref{sec:conc} concludes and discusses the limitations of our
approach.

\section{Thin-film equation for low surfactant surface coverage}
\mylab{sec:filmsurf-class}
%
\begin{figure}[t]
\includegraphics[width=0.5\hsize]{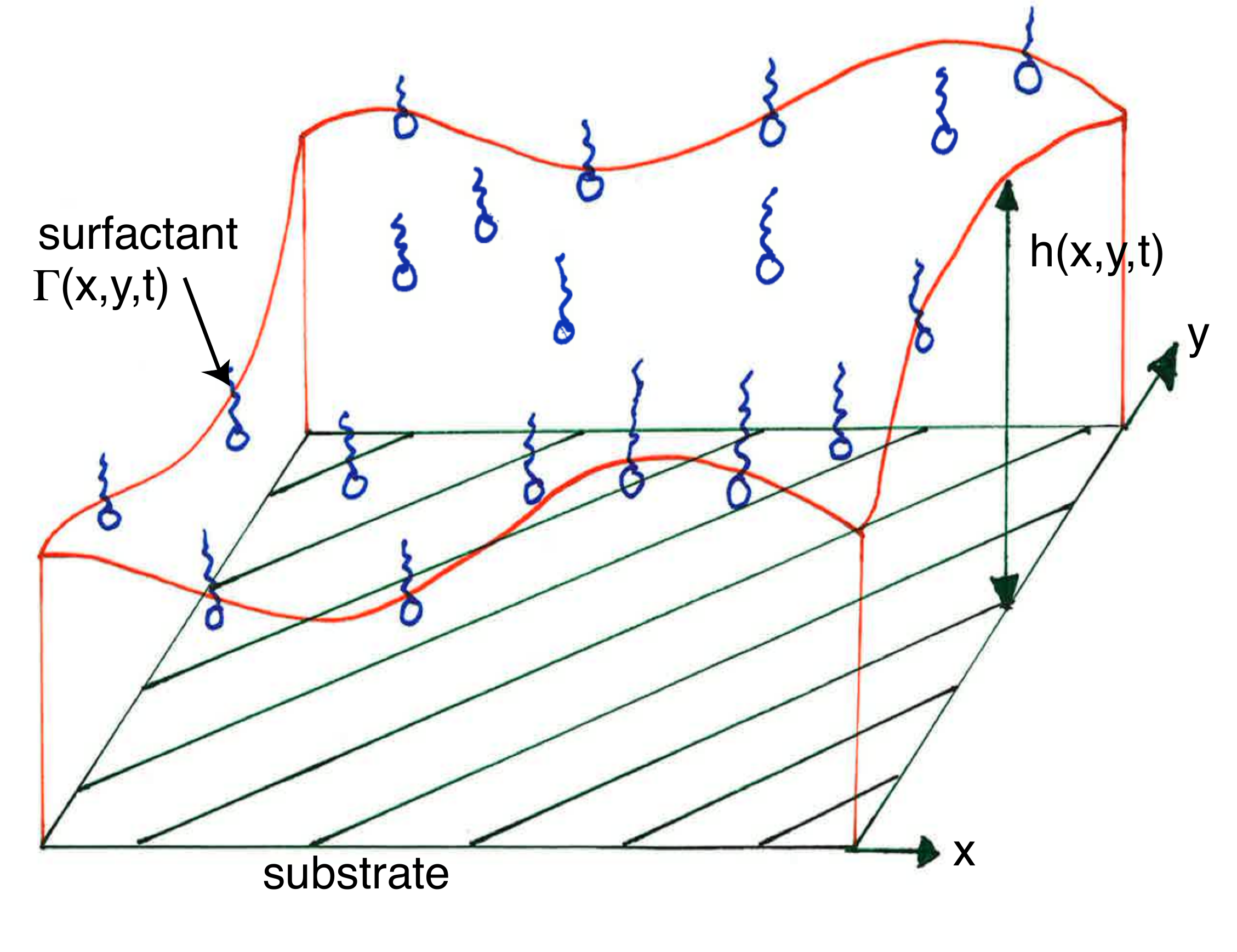}
\caption{Sketch of a surfactant covered liquid film.
}
\mylab{fig:sketch}
\end{figure}

If a hydrodynamic system involves a free surface that is covered by an
insoluble surfactant, the boundary conditions for the momentum equation
have to be supplemented by an evolution equation for the surfactant
concentration on the free surface that accounts for transport of the
surfactant by advection and diffusion and also for shape changes of
the surface that act as effective source/sink terms
\cite{Ston90,Leal07}. This equation must be solved in conjunction with the
hydrodynamic equations and boundary conditions for the liquid film.
These equations can
be greatly simplified for the case of a thin film of liquid on a solid
substrate. If all quantities in the film vary over distances with a length
scale parallel to the substrate that is large as
compared to all length scales perpendicular to it, one may make a long-wave
approximation \cite{ODB97,CrMa09} to obtain coupled evolution
equations for the film thickness profile $h(\rr,t)$ and the surfactant
surface coverage profile $\Gamma(\rr,t)$, which is a dimensionless surface
packing fraction (or concentration), where $\rr=(x,y)$ is a cartesian
coordinate over the surface. The surface concentration is defined as
$\Gamma(\rr,t) =l^2\rho (\rr,t)$, where $\rho(\rr,t)$ is the surface
number density (number per area) and $l^2$ is the
surface area per surfactant molecule when the surfactant molecules
are at maximum packing on the surface (i.e.\ $l$ is a
molecular length scale), so that close packing corresponds to
$\Gamma=1$. For the three-dimensional
physical situation illustrated in Fig.~\ref{fig:sketch}, the equation for the film
height is
\begin{equation}
\partial_t h \,=\,
- \nabla\cdot\left[\frac{h^3}{3\eta}\nabla\left(\gamma_0 \Delta h - p_\mathrm{add}(h)\right)\right] 
\,-\, \nabla\cdot \left(\frac{\gamma_\Gamma h^2}{2\eta}\nabla\Gamma\right).
\mylab{eq:govh}
\end{equation}
Note that this equation is equally valid if one uses $\rho$ instead of
the dimensionless $\Gamma$ or indeed any other measure of the surfactant surface
density. Only the quantity $\gamma_\Gamma$ needs to be redefined so that
the product of $\gamma_\Gamma$ with the surface density yields a quantity
with the dimensions of an energy per area [c.f.\ Eq.\ \eqref{eq:state-lin}].
As a result, many papers in the literature do not mention what units they
choose for $\Gamma$.
The time evolution equation for $\Gamma$ is
\begin{equation}
\partial_t \Gamma\,=\, - \nabla\cdot \left[\frac{h^2 \Gamma}{2\eta} \nabla\left(\gamma_0\Delta h - p_\mathrm{add}(h)\right)\right]
\,-\, \nabla\cdot \left(\frac{\gamma_\Gamma h\Gamma}{\eta}\nabla\Gamma\right) + 
\nabla\cdot( D \nabla \Gamma),
\mylab{eq:govgam}
\end{equation}
where $\gamma_0$ is the liquid-gas surface tension and $\eta$ is the
dynamic viscosity of the pure liquid. Partial
derivatives with respect to time and space are denoted $\partial_t$ and
$\partial_x$, respectively,  $\nabla=(\partial_x,\partial_y)$ is the
planar gradient operator and
$\Delta=\partial_{xx}+\partial_{yy}$ is the Laplace operator.
The mobility $Q(h)=h^3/3\eta$ results
from Poiseuille flow in the film without slip at the substrate. The
pressure $p=-\gamma_0 \Delta h + p_\mathrm{add}(h)$ contains the Laplace surface
curvature contribution to the pressure and additional contributions such as a
hydrostatic or a disjoining pressure
\cite{deGe85,JeGr92,DGC94,ODB97,MaCr09}.  Note that the latter is
normally assumed to be independent of $\Gamma$. Exceptions are
discussed below.  The diffusive transport of the surfactant in
Eq.~(\ref{eq:govgam}) follows from Fick's law for the flux $J_\mathrm{diff}=-D \nabla
\Gamma$. In most papers, it is assumed that the diffusion constant $D$ does
not depend on the surfactant concentration, i.e., the term
$\nabla\cdot( D \nabla \Gamma)$ in Eq.\ \eqref{eq:govgam}
becomes $D\Delta\Gamma$.

Note that Eq.~(\ref{eq:govgam}) is an equation obtained in the long-wave
approximation and therefore does not include the source-like surface
dilatation term. Different forms for such a term are discussed in
Refs.~\cite{Ston90,CFG05,PTTK07b}. For the same reason, the $\nabla$
operator in the diffusion term is the planar operator and not the operator $\nabla_s$ that acts
tangentially to the free surface. To extend the present ideas to more
general geometries, these contributions must be taken into account.

To obtain Eqs.~(\ref{eq:govh}) and (\ref{eq:govgam}), we have related
the surfactant surface coverage $\Gamma$ to the surface tension $\gamma$
by the linear equation of state in Eq.~(\ref{eq:state-lin}), i.e., a linear
solutal Marangoni effect is assumed.
In deriving Eq.~(\ref{eq:govh}), one also
assumes that $\gamma_0\gg\gamma_\Gamma (\Gamma_0-\Gamma)$ and that therefore
the Laplace pressure term ($-\gamma_0\Delta h$) only depends on $\gamma_0$.

To incorporate effects of high surfactant concentration, the equations
are often extended by translating $\gamma_\Gamma \nabla\Gamma$ back
into $\nabla\gamma(\Gamma)$, and then replacing the linear equation of
state in Eq.\ (\ref{eq:state-lin}) by some non-linear equation of state. There
are problems with doing this, as we show below.
Another extension is to incorporate a surfactant-dependent
wettability into the evolution equations (\ref{eq:govh}) and
(\ref{eq:govgam}). This is sometimes done in an ad-hoc manner by
simply replacing $p_\mathrm{add}(h)$ by some
$p_\mathrm{add}(h,\Gamma)$.  However, it turns out that this leaves the equation
incomplete and may even result in qualitatively incorrect predictions.
After re-formulating the evolution equations as a gradient dynamics in
section~\ref{sec:filmsurf-graddyn}, we discuss such extensions in
Section~\ref{sec:filmsurf-ext}. First, however, we introduce the
gradient formulation for the decoupled thin-film equation
(section~\ref{sec:pure}) and surfactant surface diffusion equation
(section~\ref{sec:diff}).
%
\section{Thin-film of pure liquid - evolution equation as a gradient dynamics}
\mylab{sec:pure}

It was noted some time ago that the time evolution equation for the film thickness in the case
without surfactant [Eq.~(\ref{eq:govh}) with $\Gamma=0$] can be written in a
variational form \cite{OrRo92,Mitl93}. This allows one to
appreciate that Eq.\ (\ref{eq:govh}) corresponds to a time
evolution equation for a conserved order parameter field $h(\rr,t)$
(cf.~Ref.~\cite{Lang92}) that follows a dissipative gradient
dynamics governed by the following equation
\begin{equation}
\partial_t h \,=\,
\nabla\cdot\left[Q_{hh}\nabla\frac{\delta F}{\delta h}\right].
\mylab{eq:varh}
\end{equation}
This equation describes how the field $h$ evolves towards a minimum of the free energy functional
\begin{equation}
F[h]\,=\,
\int\left[\gamma_0 + \frac{\gamma_0}{2}(\nabla h)^2 + f(h) \right]\,dA 
\mylab{eq:en1}
\end{equation}
where $f(h) = \int p_\mathrm{add}(h)dh$, the
mobility function $Q_{hh}=h^3/3\eta$ (c.f.~also Ref.\ \cite{Thie10})
and $dA$ is a cartesian area element
along the substrate.

Note that the free energy in Eq.\ (\ref{eq:en1}) corresponds to the one that is
obtained making a small slope
approximation in the free energy $F=\int f(h) dA + \int \gamma_0 dS$, where the surface
element $dS=\sqrt{1+(\nabla h)^2}\,dA \equiv \xi dA$ is approximated
using $\xi\approx 1+(\nabla h)^2/2$. The constant part $\int \gamma_0
dA$ of the free energy in Eq.\ (\ref{eq:en1}) is normally omitted since it does not contribute to
the dynamics, as one can see from Eq.~(\ref{eq:varh}). A similar formulation is given
in the following section for the surfactant surface diffusion equation.
%
\section{Diffusion equation as a gradient dynamics}
\mylab{sec:diff}
%

The diffusive transport of a species with small surface coverage $\Gamma$ is described by
the diffusion equation
\begin{equation}
\partial_t \Gamma \,=\,
- \nabla\cdot J_\mathrm{diff}\,=\,D\Delta\Gamma,
\mylab{eq:diff}
\end{equation}
where the diffusive flux is given by Fick's law
$J_\mathrm{diff}=-D\nabla\Gamma$. The time evolution
equation for the surfactant density in Eq.~\eqref{eq:govgam},
reduces to the diffusion equation in Eq.\ \eqref{eq:diff}
in the limit when the liquid film thickness $h(\rr,t)$ is a constant and the coefficient
$\gamma_\Gamma=0$; i.e.\ when there is no Marangoni effect.

The form in Eq~(\ref{eq:diff}) can easily obscure
the underlying thermodynamics, which can be seen when this
equation is formulated as a gradient
dynamics based on the Helmholtz free energy for an ideal gas
(i.e.\ a system of non-interacting particles):
\begin{equation}
F[\Gamma]\,=\,\frac{kT}{l^2}\int \Gamma [\log(\Gamma) - 1] \,dA,
\mylab{eq:en-diff}
\end{equation}
where $k$ is Boltzmann's constant and $T$ is the temperature.
The transport equation for $\Gamma$ is of the same form as Eq.~(\ref{eq:varh}) and
reads \cite{Evan79,MaTa99,MaTa00,ArEv04,ArRa04}:
\begin{equation}
\partial_t  \Gamma\,=\,
\nabla\cdot\left[Q_{\Gamma\Gamma}\nabla\frac{\delta F}{\delta \Gamma}\right],
\mylab{eq:trans-diff}
\end{equation}
where the mobility $Q_{\Gamma\Gamma}=\widetilde{D} \Gamma$. Here, $\widetilde{D}$ 
is the molecular mobility related to the diffusion process and may in principle depend 
on all the independent variables, although in the following we will assume that 
$\widetilde{D}$ is constant.

The equivalence of the formulation in Eq.~(\ref{eq:diff}) and in
Eqs.~\eqref{eq:en-diff} and (\ref{eq:trans-diff}) is easily established. The advantage of
the gradient dynamics form in Eq.~(\ref{eq:trans-diff}) is the ``built-in''
straightforward way to extend the description, e.g., to incorporate
attractive forces between the diffusing molecules and the effects of
higher concentrations. For instance, replacing the functional
in Eq.~(\ref{eq:en-diff}) by the one discussed by Cahn and Hilliard
\cite{CaHi58} (their equation (2.4) with (3.1)) results in the
non-linear diffusion equation (or Cahn-Hilliard equation), Eq.~(9) of Ref.\
\cite{Cahn65} (when $Q_{\Gamma\Gamma}=\widetilde{D}
\Gamma\,(1-\Gamma)$ is expanded about $\Gamma=1/2$ and only the lowest
order term is kept). More recently,
Marconi and Tarazona \cite{MaTa99,MaTa00} showed
that one can derive Eq.\ \eqref{eq:trans-diff}, starting from over-damped
stochastic equations of motion for the (surfactant) particles. They showed
that the diffusive fluid dynamics is described by Eq.~\eqref{eq:trans-diff}, taken
together with a suitable approximation for the Helmholtz free
energy functional taken from equilibrium density functional theory
\cite{Evan79,Evan92}. This so called dynamical density functional theory
\cite{MaTa99,MaTa00,ArEv04,ArRa04} is now a growing
body of work, allowing one to go beyond Cahn-Hilliard theory,
and to develop a theory which includes a microscopic
(on the scale of the particles) description of the dynamics of
particles suspended in a fluid medium, or in the present case of
surfactant particles on the surface of the liquid.

To our knowledge, no gradient dynamics formulation has yet been given
for the evolution of a thin-film covered by insoluble surfactant as
described by Eqs.~(\ref{eq:govh}) and (\ref{eq:govgam}). Since the system
is relaxational, i.e., there is no energy influx, a variational
formulation in terms of a pair of coupled evolution equations for two
conserved order parameter fields must exist and in fact is
presented in the following section.

%
\section{Evolution of a surfactant-covered film as gradient dynamics}
\mylab{sec:filmsurf-graddyn}
%

To construct a gradient dynamics description 
for the full coupled system, Eqs.~(\ref{eq:govh}) and (\ref{eq:govgam}),
we start by considering the Helmholtz free energy for the system $F[h,\Gamma]$,
that also turns out to be the Lyapunov functional for a
surfactant covered thin liquid film. It contains contributions that result
from wettability (adhesion), expressed in terms of the height profile
of the film, $\int f(h) dA$, \textit{and} contributions from the surface
$\int g(\Gamma) dS$, where $g(\Gamma)$ is the Helmholtz surface
free energy density (i.e.\ an energy per area), and $dS=\sqrt{1+(\nabla h)^2}dA=\xi
dA\approx [1+(\nabla h)^2/2]dA$ is a surface element.
The contribution of the height profile
are similar to those in Eq.~(\ref{eq:en1}). In the parameter regime
where the linear equation of state (\ref{eq:state-lin}) for the
surfactant is valid, i.e.\ when the surfactant density is low, then the
surfactant layer corresponds to a two-dimensional gas of surfactant
molecules on the film surface. The corresponding contribution to the
free energy density is the entropic (ideal-gas)
term $(kT/l^2)\,\Gamma(\log\Gamma-1)$ that on its own leads to
a diffusion equation for $\Gamma$, as discussed in section~\ref{sec:diff}. However, one may
add other contributions that are relevant at higher concentrations,
resulting from the interactions between surfactant
molecules. Such contributions are discussed below in
section~\ref{sec:filmsurf-ext}. For the non-interacting case we obtain
\begin{eqnarray}
F[h,\Gamma]\,
=\,\int\left\{f(h)+g(\Gamma)\,\xi\right\}\,dA
\mylab{eq:en2}
\end{eqnarray}
where
\begin{eqnarray}
g(\Gamma)=\,\gamma_0+ \frac{kT}{l^2}\, \Gamma [\log(\Gamma) - 1].
\mylab{eq:gg}
\end{eqnarray}
It should be noted that variations in $h$ and in $\Gamma$ are not independent:
If locally the slope of $h$ changes, the area of the liquid surface changes and 
thus the surface coverage $\Gamma$ may change without any surfactant
transport.  To derive evolution equations that are of a gradient dynamics 
form, one needs a concentration variable that is independent of the film height
profile $h$. We introduce a surface coverage $\widetilde\Gamma$, as sketched in
Fig.~\ref{fig:sketch-proj}, that corresponds to the coverage $\Gamma$ `projected' 
onto the flat substrate (thus, in principle, $\widetilde\Gamma$ can become larger 
than one). It is given by 
\begin{equation}
\widetilde\Gamma = \frac{dS}{dA}\Gamma = \xi\Gamma.
\mylab{eq:concproj}
\end{equation}

\begin{figure}[t]
\includegraphics[width=0.5\hsize]{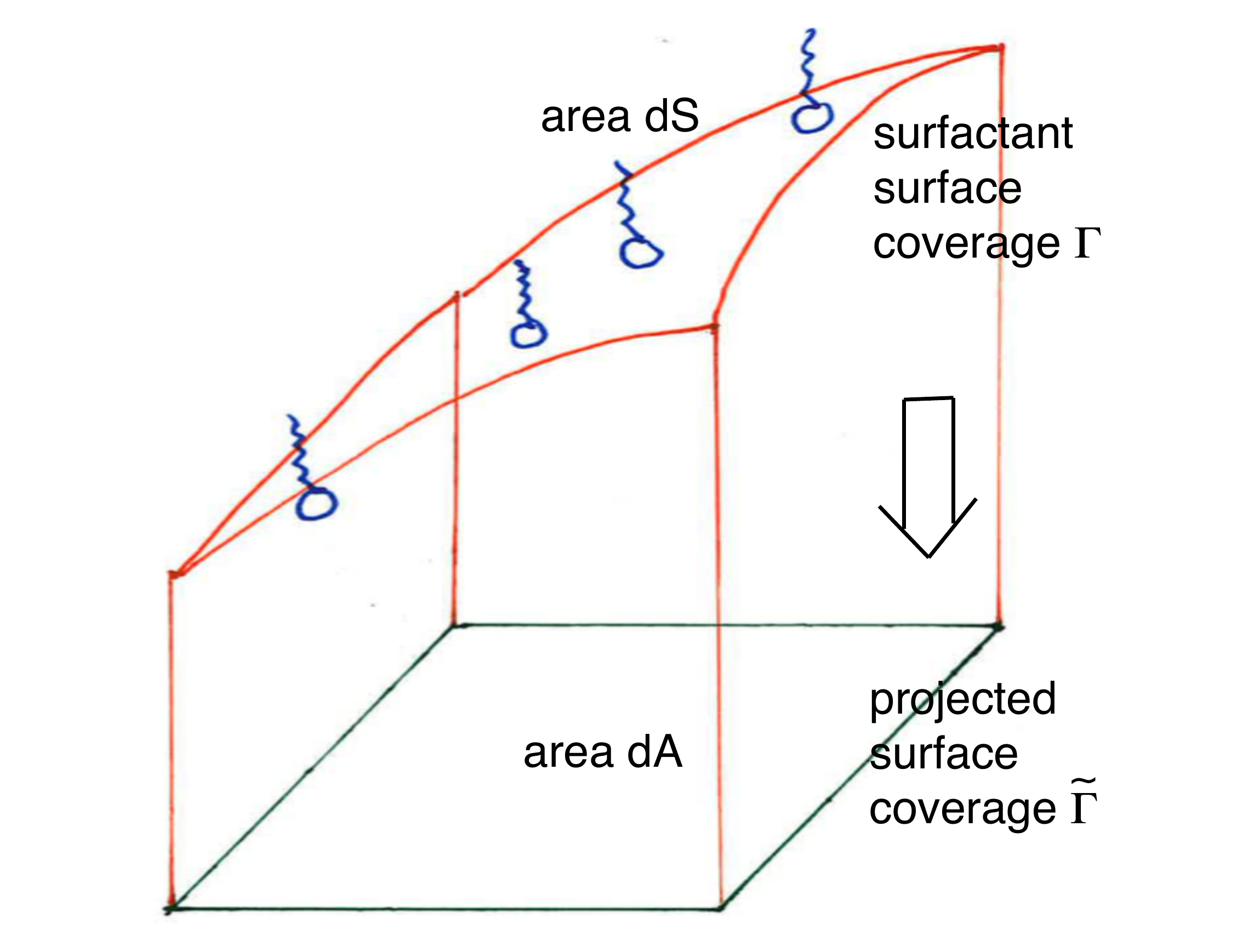}
\caption{Sketch that indicates the relation between the projected
  surface coverage $\widetilde\Gamma$ and the coverage
  $\Gamma$ on the modulated free surface, defined in Eq.~\eqref{eq:concproj}.
}
\mylab{fig:sketch-proj}
\end{figure}

Using $F[h,\widetilde\Gamma/\xi]$ from Eq.~(\ref{eq:en2}), the long-wave
hydrodynamic equations (\ref{eq:govh}) and (\ref{eq:govgam}) are
equivalent to the following general form for the time evolution equations
\begin{eqnarray}
\partial_t h \,&=&\,
\nabla\cdot\left[Q_{hh}\nabla\frac{\delta F}{\delta h}\,+\,Q_{\Gamma h}\nabla\frac{\delta F}{\delta \widetilde\Gamma}\right]
\nonumber\\
\partial_t \widetilde\Gamma \,&=&\,
\nabla\cdot\left[Q_{h \Gamma}\nabla\frac{\delta F}{\delta h}\,+\,Q_{\Gamma \Gamma}\nabla\frac{\delta F}{\delta \widetilde\Gamma}\right]
\mylab{eq:coup}
\end{eqnarray}
where the symmetric positive definite mobility matrix is
\begin{equation}
\mathbf{Q}\,=\,\left(
\begin{array}{cc}  
Q_{hh} & Q_{\Gamma h} \\[.3ex]
Q_{h \Gamma} &Q_{\Gamma \Gamma}
\end{array}
\right)
\,=\,\left( 
\begin{array}{cc}  
\frac{h^3}{3\eta} & \frac{h^2\Gamma}{2\eta} \\[.3ex]
\frac{h^2\Gamma}{2\eta} & \frac{h\Gamma^2} {\eta}+ \widetilde{D}\Gamma
\end{array}
\right).
\mylab{eq:mob}
\end{equation}
Note that we have written $\mathbf{Q}$ in terms of $\Gamma$ and $h$. The justification for
using $\Gamma$ and not $\widetilde\Gamma$ in the long-wave approximation mobilities
will become clear below. To fully appreciate the equivalence of Eq.~\eqref{eq:coup}
[with Eqs.~\eqref{eq:en2}, \eqref{eq:gg} and \eqref{eq:mob}]
to Eqs.~\eqref{eq:govh} and \eqref{eq:govgam}, we calculate the
variations of the free energy in Eq.~\eqref{eq:en2}:
\begin{eqnarray}
\frac{\delta  F}{\delta h} \,&=&\, \partial_h f(h) - \nabla\cdot \left[\omega \nabla h\right],
\nonumber\\
\frac{\delta  F}{\delta \widetilde{\Gamma}} \,&=&\,g',
\mylab{eq:insolsurf-variations}
\end{eqnarray}
where we have introduced the local surface grand potential density $\omega=g-\Gamma g'$.
Inserting these results into Eq.~\eqref{eq:coup}, we obtain the following time-evolution equations:
\begin{eqnarray}
\partial_t h \,&=&\,
\nabla\cdot\left[\frac{h^3}{3\eta}\nabla[\partial_h f(h) - \nabla\cdot
  (\omega \nabla h)]\,
+\,\frac{h^2\Gamma}{2\eta}\nabla g'\right],
\nonumber\\
\partial_t \widetilde\Gamma \,\approx\,\partial_t \Gamma&=&\,
\nabla\cdot\left[\frac{h^2\Gamma}{2\eta}\nabla\left[\partial_h f(h) - \nabla\cdot (\omega \nabla h)\right]\,+\,\left(\frac{h\Gamma^2} {\eta}+ \widetilde{D}\Gamma\right)
\nabla g' \right],
\mylab{eq:insolsurf-evolequ2}
\end{eqnarray}
where we have used the fact that $(\nabla h)^2\ll1$ to approximate $\xi\approx 1$
in the left-hand side of the second equation, as is appropriate in the long-wave
limit. Used at this stage, this approximation leads to
$\widetilde\Gamma\approx\Gamma$. Note, however, that this
approximation should be applied with caution \footnote{Note that the approximation $(\nabla h)^2\ll1$
  should only be employed in the final time-evolution equation. If one were
  to make this approximation in the free energy equation, before
  taking the functional derivative, several physically 
  important terms end up being omitted from the final time-evolution
  equation.}.
The equations for $\partial_t h$ and $\partial_t \Gamma$
exactly correspond to the hydrodynamic model [Eqs.~(\ref{eq:govh}) and
(\ref{eq:govgam})] if (i) one identifies the local surface grand potential
density $\omega$ with the surface tension $\gamma$ (see further discussion in
Section~\ref{sec:eqstate} below) and (ii) employs Eq.~(\ref{eq:gg}) for
$g(\Gamma)$, i.e., one only includes the ideal-gas contribution to the free energy, which is
a reasonable approximation to make for low concentrations of surfactant. As a result 
\begin{equation}
\nabla \gamma = - \Gamma \nabla \delta F /
\delta\tilde\Gamma=\Gamma\nabla g'= (kT/l^2)\nabla\Gamma,
\mylab{eq:insolsurf-gamrel}
\end{equation}
and so one finds that the
diffusion coefficient $D$ and the solutal Marangoni coefficient
$\gamma_\Gamma$ in Eqs.~(\ref{eq:govh}) and (\ref{eq:govgam}) are
given by $D=kT \widetilde{D}/l^2$ and $\gamma_\Gamma=-kT/l^2$,
respectively.

Note that the hydrodynamic community often assumes
that the change of the surface tension with concentration is small as
compared to the reference surface tension $\gamma_0$ and therefore
only uses $\gamma_0$ in the Laplace pressure term, i.e., $\nabla\cdot
(\omega \nabla h)\approx \gamma_0\Delta h$.

\subsection{Equation of state - surface tension}
\mylab{sec:eqstate}
%

Before presenting several extensions to the hydrodynamic equations
(\ref{eq:govh}) and (\ref{eq:govgam}), based on our reformulation in Eq.\
(\ref{eq:insolsurf-evolequ2}), we put our gradient dynamics
formulation in its proper thermodynamic context.
First, we discuss the equation of state that relates surface tension
$\gamma$ to surfactant concentration $\Gamma$,
and show that the thermodynamic and hydrodynamic approaches are fully
consistent for any dependence $g(\Gamma)$. 

To this end, we review first some elementary considerations
concerning the relation between the surface tension and the surface free
energy density. The latter is defined as the excess free energy per unit area that is due 
to the presence of a surface. In analogy to bulk thermodynamics, this excess
may be defined for different thermodynamic ensembles and may therefore
depend on different surface thermodynamic variables. Moreover, this surface excess
acts as the thermodynamic potential for surface variables (for an
enlightening general discussion of surface excesses, see 
Ref. \cite{Cahn79}). The surface tension is the
derivative of this surface thermodynamic potential with respect to the area --
it is the equivalent of the pressure in bulk thermodynamics 
(up to a sign convention: a positive pressure generates 
an outward force on the container walls, whereas a positive 
surface tension creates an inward force on the lines bordering 
a surface element).

It is easy to show that the surface tension is always given by the
surface excess grand potential density, regardless of whether the
surface free energy is defined in the canonical or grand-canonical
ensemble. In the canonical case, the surface free energy is the surface (excess)
contribution to the Helmholtz free energy, $F_{\rm surf}=Sg(\Gamma)$, for a surface 
element of area $S$. The variation of this free energy is 
$dF_{\rm surf}=g(\Gamma)dS + Sg'(\Gamma)d\Gamma$. The second term 
arises from the fact that the variation in the canonical ensemble has
to be taken for a fixed number of surfactant molecules, $N=S\Gamma/l^2$ 
(recall that $\Gamma=\rho l^2$), and thus the variation of the surface 
area creates a variation in the local concentration equal to
$d\Gamma=d(l^2N/S)=-(l^2N/S^2)dS=-(\Gamma/S)dS$, and the surface tension becomes
$\gamma=\omega(\Gamma)=dF_{\rm surf}/dS=g(\Gamma)-g'(\Gamma)\Gamma$,
where we have introduced the surface grand potential density
$\omega(\Gamma)=g(\Gamma)-\mu\Gamma$, with $\mu=g'(\Gamma)=\partial g/\partial
\Gamma$ the chemical potential \footnote{Note that here we define the
chemical potential $\mu=\partial g/\partial \Gamma$, although strictly it is defined as
the derivative $\partial g/\partial \rho$. This definition,
which differs from the usual one by a factor of $l^2$, leads to a simplification
of some of our equations.}. Alternatively, if the surface
excess is defined in the grand-canonical ensemble, the surface free energy
is directly given by $F_{\rm surf}=S\omega$. Now, the variation has to be taken
at constant chemical potential (the surface element is connected
to a reservoir of surfactant molecules). Since $\Gamma$ is a function of the 
chemical potential in the grand-canonical ensemble, it remains fixed, and therefore
the surface tension is directly given by $\gamma=dF_{\rm surf}/dS=\omega$
\footnote{Note that formally, $\omega$ is not
a function of $\Gamma$, but of $\mu$, so when we write $\omega(\Gamma)$,
this should be read as $\omega(\Gamma(\mu))$.}.

For flat interfaces with a small surfactant concentration, the
surface-related part of the local Helmholtz free energy $g$ is given by
Eq.~(\ref{eq:gg}), and so the chemical potential
$\mu=(kT/l^2)\,\log\Gamma$. With this, the surface tension becomes
\begin{eqnarray}
\gamma &=& \omega = g(\Gamma)-\mu\Gamma\\
&=&\gamma_0 - \frac{kT}{l^2}\Gamma,
\mylab{eq:eos2b}
\end{eqnarray}
i.e., one recovers the linear dependence, Eq.~(\ref{eq:state-lin}), used in
hydrodynamics with
$\gamma_\Gamma=-kT/l^2$ and $\Gamma_0=0$.
Note that if the surface tension is defined (incorrectly)
as the local Helmholtz free energy $g(\Gamma)$, the logarithmic terms
entail that one does not recover the linear dependence in Eq.~(\ref{eq:state-lin}).

By identifying the surface tension with the local
grand potential, the thermodynamic and hydrodynamic formulations are
fully consistent for any convex local $g(\Gamma)$.
In the hydrodynamic
formulation [Eqs.~(\ref{eq:govh}) and (\ref{eq:govgam})] the Marangoni
force contributes to the advective flux as
$-(h^2/2\eta)\,\nabla\gamma(\Gamma)$, and normally the equation of
state $\gamma(\Gamma)$ is given directly. With
$\gamma=\omega=g-\mu\Gamma$, the Marangoni term becomes
$(h^2\Gamma/2\eta)\,(\partial_{\Gamma\Gamma}g)\,\nabla\Gamma$. This
expression is identical to the term one obtains in the variational
formulation~(\ref{eq:coup}), i.e., $Q_{\Gamma h}\nabla(\delta
F/\delta\tilde{\Gamma})=(h^2\Gamma/2\eta)\,\nabla(\delta F/\delta\tilde{\Gamma})$. For
any $F$ of the form in Eq.~(\ref{eq:en2}), this equals
$(h^2\Gamma/2\eta)\,(\partial_{\Gamma\Gamma}g)\,\nabla\Gamma$.  The
equivalence of the Marangoni term in the hydrodynamic and thermodynamic
formulation is valid for any convex local $g(\Gamma)$. In other words, the surface
tension gradient $\nabla\gamma$ may be expressed either as
$\nabla\omega$ or as $-\Gamma\nabla g'$. This implies that
Eq.~(\ref{eq:insolsurf-evolequ2}) may be written as 
\begin{eqnarray}
\partial_t h \,&=&\,
\nabla\cdot\left[\frac{h^3}{3\eta}\nabla[\partial_h f(h) - \nabla\cdot
  (\gamma \nabla h)]\,
-\,\frac{h^2}{2\eta}\nabla\gamma\right]
\nonumber\\
\partial_t \Gamma&=&\,
\nabla\cdot\left[\frac{h^2\Gamma}{2\eta}\nabla\left[\partial_h f(h) - \nabla\cdot (\gamma \nabla h)\right]
\,-\,\left(\frac{h\Gamma} {\eta}+ \widetilde{D}\right)
\nabla \gamma \right].
\mylab{eq:insolsurf-evolequ3}
\end{eqnarray}
Thus, the diffusion term is expressed in terms of $\nabla\gamma$
and the molecular mobility $\widetilde{D}$. Note, however, that this
argument no longer holds if the free energy functional contains
non-local terms in $\Gamma$, such as, e.g., $(\nabla\Gamma)^2$. Then the
formulation in Eq.~(\ref{eq:insolsurf-evolequ3}) can not be used and one
must start directly with Eqs.~(\ref{eq:coup}).
%
\section{Extensions}
\mylab{sec:filmsurf-ext}
%

Up to this point, we have presented a gradient dynamics re-formulation of the
hydrodynamic long-wave model for the evolution of a thin-film that is
covered by a low concentration of insoluble surfactants, that has
several connections to the approach taken in dynamical density functional theory,
with a local approximation for the free energy.

The present re-formulation really demonstrates its advantages when seeking to make
(common) extensions of the hydrodynamic model, such as to incorporate nonlinear
equations of state, surfactant-dependent wettability, phase
transitions at high surfactant concentrations, or substrate mediated
surfactant condensation. Such effects are often included into the
hydrodynamic formulation (\ref{eq:govh})
and (\ref{eq:govgam}) in an \textit{ad hoc} manner that may result in
the omission of important terms and sometimes lead to qualitatively incorrect
behaviour. The thermodynamic variational framework presented here, i.e., 
Eqs.~(\ref{eq:insolsurf-evolequ2}), allows us to make extensions
stemming from changes (extra terms) in the free energy functional (\ref{eq:en2}) in a
systematic and thermodynamically consistent manner.  In the following, we discuss
several examples.

\subsection{Nonlinear equation of state}

The most common extension is to replace the linear
equation of state in Eq.~(\ref{eq:state-lin}) by various nonlinear
expressions.  Examples include the exponential (Gaussian) dependence
$\gamma=C_1+C_2\exp(-\Gamma^2/C_3)$ \cite{SHZD11}; a smoothed
step-wise change $\gamma=C_1+C_2\tanh[C_3(\Gamma-1)]$
\cite{LoHi00,Heid02,LFB06}; Langmuir-Szyszkowski
($\gamma=C_1+C_2\log(1-\Gamma)$) \cite{PaSt96,JaLo04,KvM04,LFB06,Hame08} or
Frumkin ($\gamma=C_1+C_2\log(1-\Gamma)+C_3\Gamma^2$)
\cite{PaSt96,ChBo05,LFB06} equations applied to insoluble surfactants;
expressions related to power laws, such as, e.g.,
$\gamma=C_1+C_2(1+C_3\Gamma)^{-3}$
\cite{BoGr88,GaGr90,GaGr92,JeGr92,WCM04b}); and fits to experimentally
obtained isotherms \cite{Bull99}. In all cases the $C$'s
represent various constants.
In most works, the extension is done by solely replacing the parameter
$\gamma_\Gamma$ in Eqs.~(\ref{eq:govh}) and (\ref{eq:govgam})
by the function
$\gamma' =\partial\gamma(\Gamma)/\partial \Gamma$.  This, however,
does not take into account that in addition to the convective transport due to
the Marangoni force, the diffusive transport of the surfactant is also
affected when the underlying free energy functional changes. Thus, most
works assume that the diffusion constant $D$ remains independent of
the surfactant concentration, even when working with highly nonlinear
equations of state.  An exception is Ref.~\cite{BoGr88} that uses a
$D(\Gamma)$. 

Based on our thermodynamic reformulation,
the proper relation between $\gamma$ and $D$ (discussed
at the end of section~\ref{sec:eqstate}) results in the amended
hydrodynamic equations~(\ref{eq:insolsurf-evolequ3}) that are expressed
in terms of $\gamma(\Gamma)$. One may define a non-constant
$D(\Gamma)$ by enforcing Ficks's law
\begin{equation}
J_\mathrm{diff}=-D\nabla\Gamma
\label{eq:Fick_J}
\end{equation}
to hold. From a comparison of 
Eqs.~(\ref{eq:govgam}) and (\ref{eq:insolsurf-evolequ2}) one obtains
$D\nabla\Gamma=\widetilde{D}\Gamma\nabla g'=\widetilde{D}\Gamma
g''\nabla\Gamma$, i.e., $D=\widetilde{D}\Gamma g''$. With this $D$ the
hydrodynamic formulation (\ref{eq:govh}) and (\ref{eq:govgam}) is
consistent with the gradient dynamics form
(\ref{eq:insolsurf-evolequ2}). To obtain $D$ in terms of the equation
of state, one differentiates the relation $\gamma=g-\Gamma g'$ with respect to
$\Gamma$. The result $\gamma'=-\Gamma g''$ implies
\begin{equation}
D=-\widetilde{D}\gamma'.
\mylab{eq:dst}
\end{equation}
Note, however, that in principle $\widetilde{D}$ itself might also be a
function of $\Gamma$ (and other state variables). The relations
$\gamma(\Gamma)$ and $D(\Gamma)$ employed in Ref.~\cite{BoGr88} are
only consistent with the general thermodynamic framework given here,
if particular dependencies of $\widetilde{D}$ on $\Gamma$ are assumed.
Note that Fick's law in Eq.~\eqref{eq:Fick_J} is only
true in the low density $\Gamma\to0$ limit. More generally one should have
\begin{equation}
J_{\rm diff}= -M\nabla \mu
\end{equation}
where $M(\Gamma)$ is a mobility coefficient \cite{Darken48,MaTa99,MaTa00,ArEv04,ArRa04}.
This form is universally valid,
whereas Eq.\ \eqref{eq:Fick_J} does not always hold (e.g.\ in the `uphill diffusion'
observed in spinodal decomposition).

Note also that many authors correctly employ the hydrodynamic form
in Eqs.\ (\ref{eq:govgam}) and (\ref{eq:insolsurf-evolequ2}) with a linear
equation of state, but give the Marangoni flux in its general form as
$J_\mathrm{mar} =(h^2/2\eta)\nabla\gamma(\Gamma)$ in combination with
a surfactant-independent diffusion constant
\cite{JeGr93,MaTr99,WCM02b,CrMa09,PeSh11}. Whilst this approach
is indeed correct for a linear equation of state, it should be stressed that
it is not valid for arbitrary (nonlinear) equations of state.

\subsection{Surfactant-dependent wettability}

It is widely accepted that wettability depends on the surface density of the
surfactants \cite{Chur95,BhRu97}.  However, the literature is less clear on how
such effects may be incorporated in a hydrodynamic thin-film
description by extending the model given in Eqs.~(\ref{eq:govgam}) and
(\ref{eq:insolsurf-evolequ2}). In all the contributions we are aware
of, this is done by replacing the film-thickness dependent
Derjaguin (or disjoining) pressure $\Pi(h)$, that is contained in the
pressure $p_\mathrm{add}(h)$ in Eqs.~(\ref{eq:govh}) and
(\ref{eq:govgam}), by a disjoining pressure $\Pi(h,\Gamma)$ that
depends on film thickness and surfactant concentration
\cite{WCM02,Hu05,FiGo07,CrMa07}. The influence of surfactants on
the various components of Derjaguin's pressure for thin-films are
 discussed in detail in Refs.~\cite{BhRu97,DIAL06} in the context of free
standing (soap) films. For a simple model for forces between surfaces
with adsorbed layers see Ref.~\cite{Isra11}.

Based on our thermodynamic re-formulation, one can now see how
the hydrodynamic equations~(\ref{eq:govh}) and
(\ref{eq:govgam}) must be amended to account for any dependency of
the adhesion energy on film thickness and surfactant concentration. Replacing
$f(h)$ in the free energy functional~(\ref{eq:en2}) by $f(h,\Gamma)$ results in the
additional contributions $\partial_\Gamma f$ to $\delta F/\delta
\Gamma$ that affect both evolution equations. The resulting
hydrodynamic form is
\begin{eqnarray}
\partial_t h \,&=&\,
\nabla\cdot\left[\frac{h^3}{3\eta}\nabla[\partial_h f(h,\Gamma) - \nabla\cdot
  (\gamma \nabla h)]\,
-\,\frac{h^2}{2\eta}[\nabla \gamma -\,\Gamma\nabla \partial_\Gamma f(h,\Gamma)] \right]
\nonumber\\
\partial_t \Gamma&=&\,
\nabla\cdot\left[\frac{h^2\Gamma}{2\eta}\nabla\left[\partial_h f(h,\Gamma) -  \nabla\cdot (\gamma \nabla h)\right]
\,-\,\left(\frac{h\Gamma} {\eta}+ \widetilde{D}\right)
[\nabla \gamma -\,\Gamma\nabla \partial_\Gamma f(h,\Gamma)]
\right].
\mylab{eq:insolsurf-evolequ4}
\end{eqnarray}
The extra terms should be interpreted as an additional contribution to the
Marangoni force that must be taken into account for small film
thicknesses. The effective Marangoni force is
$\nabla\gamma+\Gamma\nabla \partial_\Gamma f(h,\Gamma)$. It also becomes
the effective driving force for diffusion of the surfactant. To our knowledge these
terms have not been included in any of the thin-film evolution
equations that model surfactant-covered ultrathin films. However, they
are necessary in any model that involves a surfactant-dependent
Derjaguin pressure. Without them, the model may exhibit qualitatively
incorrect behaviour, such as oscillatory instability modes \cite{FiGo07}, as discussed in
more detail below. Furthermore, as the system evolves
in time, it does not tend to the correct equilibrium state, particularly
in the contact line region.
In short, to treat such effects correctly one must determine a
suitable form for $f(h,\Gamma)$, obtained from the surfactant-dependent
Derjaguin pressures discussed in the literature
\cite{Chur95,BhRu97,DIAL06}.

\subsection{Surfactant phase transitions}

A third example is the description of phase separating surfactant
mixtures or phase transitions at high surfactant concentrations. The
simplest case of surfactant molecules that slightly attract each other
is already addressed by the discussion above, as it only results in a nonlinear
equation of state. For instance, for weakly attracting surfactant molecules, one
replaces the purely entropic form in Eq.~(\ref{eq:gg}) by
$g(\Gamma)=\gamma_0+ \frac{kT}{l^2}\, \Gamma [\log(\Gamma) - 1]
-(a/2)\Gamma^2$, where the attraction strength parameter $a>0$. This results in
$\gamma=\gamma_0 - (kT/l^2)\Gamma + (a/2)\Gamma^2$ and so the
effective diffusion constant depends linearly on $\Gamma$.

The situation becomes more involved for surfactant layers that can
undergo a phase transition when the concentration changes, e.g.,
between the gaseous and the liquid-expanded or between the
liquid-expanded and the liquid-condensed phases
\cite{RuLi98,AdGa97}.
Beside a function $g(\Gamma)$ that
accounts for the particular surfactant isotherm, one also needs to
incorporate a surface gradient term $(\kappa/2)(\nabla_s \Gamma)^2\xi$
in the free energy functional~(\ref{eq:en2}) to account for the finite
width and line tension of the interface between the various surfactant
phases. If a double-well
potential is used for $g$, this amounts to a description of the
surfactant layer using a convective Cahn-Hilliard-type equation. A
similar approach is employed in Ref.~\cite{KGF09} to describe a
thin liquid film covered with an insoluble surfactant in the vicinity
of a first-order phase transition. However, as explained below, our
formulation differs on a number of important points.

As the algebra is involved, we illustrate this for the one-dimensional
case, where $x$ is the only spatial coordinate. 
The free energy functional is
\begin{equation}
F[h,\widetilde\Gamma/\xi]\,
=\,\int\left\{f(h)+g\left(\frac{\widetilde\Gamma}{\xi}\right)\,\xi +
  \frac{\kappa}{2} \left(\partial_x\frac{\widetilde\Gamma}{\xi}\right)^2 \frac{1}{\xi} \right\}\,dx
\mylab{eq:en-nonlocal}
\end{equation}
where we use $\partial_s = (1/\xi)\partial_x$,
$\widetilde\Gamma=\xi\Gamma$, $ds/dx=\xi$ and $s$ is the arc-length
coordinate along the free surface.  Note that the final contribution to
the integral in Eq.~\eqref{eq:en-nonlocal} is simply the term $(\kappa/2)\int
(\partial_s\Gamma)^2ds$. We keep
$\xi=\sqrt{1+(\partial_x h)^2}$ exact throughout the derivation and
only use $(\partial_x h)^2\ll1$ at the end. Employing the
approximation too early can lead to neglecting physically essential terms,
such as the Laplace pressure.
Details of the calculation of the functional derivatives are given in
Appendix~\ref{sec:app-vari}. The resulting expressions are
\begin{eqnarray}
\frac{\delta  F}{\delta h} \,&=&\, f'  - \partial_x\left[\left(\omega -\frac{\kappa}{2}\left(\partial_x \Gamma\right)^2
+ \kappa\Gamma\partial_{xx}\Gamma
\right)\partial_x h
\right]
\nonumber\\
\frac{\delta  F}{\delta \widetilde{\Gamma}}
\,&=&\, g'  -\kappa\partial_{xx}\Gamma
\mylab{eq:nonlocal-variations}
\end{eqnarray}
where we have used $(\partial_x h)^2\ll1$ [see
appendix~\ref{sec:app-vari} -- in particular Eqs.~(\ref{eq:app-vari2})
and (\ref{eq:app-vari3})]
and $\omega=g-\Gamma g'$. The time evolution equations for $h$ and $\Gamma$
are obtained by substituting Eqs.~(\ref{eq:nonlocal-variations}) into
Eqs.~(\ref{eq:coup}). On inspecting the resulting equations, one notices
that we have again obtained the form in Eq.\ (\ref{eq:insolsurf-evolequ3}),
but now the surface tension is
\begin{equation}
\gamma=\widetilde\omega\equiv\omega -\frac{\kappa}{2}
(\partial_x\Gamma)^2 + \kappa \Gamma \partial_{xx} \Gamma.
\mylab{eq:surf-tens-pt}
\end{equation}
Recall that above, for the case without the gradient terms in $\Gamma$,
we had that $\partial_x \gamma = - \Gamma \partial_x \delta F / \delta\tilde\Gamma$
[Eq.~(\ref{eq:insolsurf-gamrel})]. It turns out that in the
present case this result still holds.
The surface grand potential density for the nonlocal case is
$\widetilde\omega=g+(\kappa/2)(\partial_x\Gamma)^2 - \Gamma \delta F / \delta\tilde\Gamma$.
This observation implies that with the proper definition of surface tension,
the evolution equations in Eq.~(\ref{eq:insolsurf-evolequ3}) are valid for both the
extension to include nonlinear equations of state and the
present extension that incorporates gradient terms in $\Gamma$ in the
free energy.

This issue explains the differences between
our formulation and that in Ref.~\cite{KGF09}, that starts from a
hydrodynamic formulation somewhat similar to that in Eq.\
(\ref{eq:insolsurf-evolequ3}) \footnote{%
Note in connection with this, one can see that there is a sign error in Ref.~\cite{KGF09},
due to the use of $(\kappa/2)\int (\partial_x\Gamma)^2ds$ instead of $(\kappa/2)\int
(\partial_s\Gamma)^2ds$ for the gradient term. This leads to the factor $1/\xi$ in
the last term of Eq.~(\ref{eq:en-nonlocal})
becoming $\xi$, resulting in the opposite sign of the
$(\nabla\Gamma)^2$ term in the prefactor of $\partial_x h$ in
Eq.~(\ref{eq:nonlocal-variations}), when the $(\partial_xh)^2\ll1$ approximation is used.
Furthermore, they use $\widetilde\omega\partial_{xx} h$ instead of
$\partial_x(\widetilde\omega\partial_{x} h)$. Their formulation
corresponds to the use of different approximations for $\Gamma$ in
the various instances it appears in the mobility
matrix in Eq.~(\ref{eq:mob}).}. 
We expect the formulation
presented here to be useful for studying the dynamics
of surfactant phase transitions on thin films, for the case of insoluble surfactants.
For instance, incorporating gradient
terms may enable one to explain the spatially non-monotonic
distribution of a spreading surfactant drop that has been observed
in recent experiments \cite{FLFD10}.

\subsection{Substrate-mediated phase transitions}

As final example, we mention the so-called substrate-mediated phase
transitions of surfactant layers that may occur when surfactant
monolayers are transferred from a deep trough onto a solid substrate,
i.e., during a Langmuir Blodgett transfer \cite{RiSp92,LBM00}. Often,
the substrate triggers a phase transition from the liquid-expanded
(LE) phase to the liquid-condensed (LC) phase of the surfactant
monolayer. Within the framework presented here, this transition may be
described by replacing the surfactant contribution to the free energy
(\ref{eq:en2}), $g(\Gamma)$, by a term that depends on both the
surfactant concentration and film height, $g(h,\Gamma)$. Doing this,
one obtains additional contributions to the free energy variations:
The $g'$ has to be replaced by $\partial_\Gamma g$ and a term
$\partial_h g$ is added to $\delta F/\delta h$ in
Eq.~(\ref{eq:insolsurf-variations}).  For the full expressions, see
appendix~\ref{sec:app-vari}.  Such a surfactant concentration and 
film-height dependent contribution to the free energy is employed in
Refs.~\cite{KGFC10,KGFT12} to describe substrate-mediated condensation,
but without incorporating the additional $\partial_h g$ term. For
their choice of $g(h,\Gamma)$, the omission is of no major
consequence; it only amounts to a redefinition of the parameters in the
disjoining pressure.

\section{Consequences of the gradient dynamics formulation}
\mylab{sec:filmsurf-consequ}

The advantage of the gradient dynamics formulation, besides its
thermodynamic consistency, is that one may readily use general results
obtained for other systems having governing equations of the form of Eq.~(\ref{eq:coup}).  Similar
formulations exists, for instance, for two-layer thin-film systems
(where the two conserved fields are the two film thicknesses)
\cite{PBMT04,PBMT05} and for thin-films of solutions or suspensions
(where the two conserved fields are the film thickness $h$ and the
effective solute layer thickness $\psi=h\phi$, where $\phi$ is the
vertically averaged concentration \cite{Thie11b}).

\subsection{Lyapunov functional}

Just as in the above-mentioned cases, one can show that the free
energy functional $F[h,\Gamma]$ in Eq.~\eqref{eq:coup} is
a Lyapunov functional: The total time derivative of $F[h,\Gamma]$ is 
$dF/dt = \int \left( \frac{\delta F}{\delta h} \partial_ t h + 
\frac{\delta F}{\delta \widetilde\Gamma} \partial_ t \Gamma \right) \,dS $.
Expressing the partial derivatives $\partial_t h$ and $\partial_t \Gamma$ by
the expressions in Eq.~(\ref{eq:coup}) and
after integration by parts and assuming periodic or no-flux boundary conditions, one obtains
\begin{equation}
\frac{dF}{dt} = - \int \left[ 
Q_{hh} \left( \nabla \frac{\delta F}{\delta h} \right)^2\,+\,
2\,Q_{h\Gamma} \left( \nabla \frac{\delta F}{\delta h} \right)
\cdot \left(\nabla \frac{\delta F}{\delta \widetilde\Gamma} \right) \,+\,
Q_{\Gamma\Gamma} \left( \nabla \frac{\delta F}{\delta \widetilde\Gamma} \right)^2
\right]\,dx. 
\mylab{LYAP_dt}
\end{equation}
Because [cf.~Eq.~\eqref{eq:mob}]
\begin{equation}
\det \vec{Q} =  \frac{h^4\Gamma^2}{12\eta^2} + \frac{\widetilde{D}h^3\Gamma}{3\eta}> 0,
\mylab{DET}
\end{equation}
and $Q_{hh} > 0$ and $Q_{\Gamma\Gamma} >0$, the quadratic
form in Eq.~(\ref{LYAP_dt}) is positive definite and therefore $dF/dt <0$,
and $F$ is a proper Lyapunov functional.  Furthermore, one may
identify the stationary solutions of Eqs.\,(\ref{eq:coup}) with the
extrema of $F$.

\subsection{Stability of flat films}

Next, we briefly discuss a general result for the linear stability of
flat films $h(x,t=0)=h_0$, that are covered by a homogeneous layer of surfactant
$\Gamma(x,t=0)=\Gamma_0$. On a system of infinite size, one may
decompose any fluctuation disturbances of the film height and surfactant concentration
in this homogeneous state into Fourier modes
and consider their time evolution. We employ the ansatz
$h(x,t)=h_0\left[1+\epsilon \exp{(\beta t+kx)}\right]$ and
$\Gamma(x,t)=\Gamma_0\left[1+\epsilon\chi \exp{(\beta t+kx)}\right]$ where $k$ and
$\beta(k)$ are the wave number and growth rate of the harmonic mode,
respectively. The overall amplitude of the disturbance is $\epsilon$,
while $\chi$ is the amplitude ratio of the disturbances in the
surfactant concentration and film thickness profiles. In short, the amplitudes
may be written in vector notation as $\epsilon\vecg{\chi}=\epsilon
(h_0,\chi\Gamma_0)^T$.

Employing these ansatzes for $h(x,t)$ and $\Gamma(x,t)$ in 
Eqs.\,(\ref{eq:coup}), and then linearising in
$\epsilon\ll1$, as is appropriate for small amplitude disturbances,
leads to the following eigenvalue problem 
\begin{equation}
(\vec{J} -\beta \vec{I}) \vecg{\chi} =0,
\mylab{eq:evprob}
\end{equation}
where $\vec{J}$ is the $non$-symmetric Jacobian given by
\begin{equation}
\vec{J} = -k^2 \vec{Q}_0\vec{E}_0
\mylab{eq:jacobian}
\end{equation}
and where $\vec{E}_0$ and $\vec{Q}_0$ are the matrix of the second
variations of $F$ in Fourier space and the mobility matrix,
respectively, both evaluated at $h_0$ and $\Gamma_0$.
Since $\det \vec{Q} \neq 0$ for $h,\Gamma>0$, Eq.~\eqref{eq:evprob}
can be written as the generalised eigenvalue problem 
\begin{equation}
(k^2\vec{E}_0 + \beta \vec{Q}^{-1}_0) \vecg{\chi} = 0.
\mylab{eq:evprob-general}
\end{equation}
Because $\vec{E}_0$ and $\vec{Q}^{-1}_0$ are both symmetric and
$\vec{Q}^{-1}_0$ is positive definite, one can deduce that all
eigenvalues $\beta$ are real \cite{MSV87}, as one should expect for a
variational problem.  Inspecting Eq.~\eqref{eq:evprob-general} further
indicates that the stability of the system is completely determined by the
eigenvalues of $\vec{E}_0$, i.e., by the second variations of the
free energy functional. The stability threshold is given by $\det
\vec{E}_0=0$. However, having $\vec{E}_0$ is not sufficient to obtain the
actual growth rate $\beta(k)$ of the unstable modes (i.e.\ the dispersion relation)
and the amplitude ratio $\chi$. These are obtained by
solving Eq.~\eqref{eq:evprob}. A remarkable effect that arises from the
coupling of the two fields, i.e., when $\delta^2 F/\delta\Gamma\delta
h\neq 0$, is that the system becomes unstable for a larger
range of parameter values than the individual (decoupled) systems are on their
own. This effect is discussed in many other contexts -- see
e.g.~\cite{FiDi97,PBMT04,Clar04}.

The fact that the present system has a non-diagonal mobility matrix $\vec{Q}$
in Eq.~\eqref{eq:mob} (as do the models in Refs.~\cite{PBMT04,Thie11b})
distinguishes it from many other systems with evolution equations for two coupled
order parameter fields having a gradient
dynamics. This means that the equations for both fields depend on
variations of the free energy with respect to each of the fields $h$ and
$\Gamma$. Thus,
for a non-diagonal $\vec{Q}$ and when $\delta^2 F/\delta\Gamma\delta h\neq 0$,
the evolution of the two fields are coupled both through the free
energy functional and through the dynamical mobility coefficients in $\vec{Q}$.
In contrast, many such systems have a diagonal mobility matrix, and then the
evolution of the two fields is solely coupled by the off-diagonal term in the
matrix of second derivatives, $\delta^2 F/\delta\Gamma\delta h$. 
Examples are the equations in Ref.~\cite{PlGo97b}, that
describe the spinodal decomposition in ternary systems,
the dewetting of nanoparticle suspensions in
Ref.~\cite{RAT11}, the coupled demixing and dewetting of a binary
mixture discussed in Ref.~\cite{Clar05}, the electric field driven surface
instability of two air-gap separated polymer layers in a capacitor
\cite{Amar12}, and the model equations employed in Ref.~\cite{FiDi97}
to describe the interplay between ordering and spinodal decomposition
in binary systems.

\section{Conclusions}
\mylab{sec:conc}

We have proposed several amendments and extensions
for models describing the dynamics of liquid films that are covered
by high concentrations of insoluble surfactant. After briefly reviewing the
`classical' hydrodynamic form of the coupled evolution equations for
the film height and the surfactant concentration profile, that are well established for
small concentrations, we have re-formulated the model in three stages as a
gradient dynamics. We refer to this as the ``thermodynamic form'' of
the evolution equations.

In the first stage, we have given the gradient dynamics form of the
evolution equation for a thin film of a pure liquid on a flat substrate
without surfactant, in the case where capillarity and wettability are
the dominant influences. This formulation was discussed before, e.g.\ in
Refs.~\cite{Mitl93,Thie10}, and is of the standard form suitable for conserved
dynamics, of which a classic example is the Cahn-Hilliard equation \cite{Cahn65}
for the demixing dynamics of a binary mixture. In the second stage, we have
briefly reviewed the classical diffusion equation and have noted the
existence of a gradient dynamics formulation that puts it in the context of
nonequilibrium thermodynamics and dynamical density functional
theory.  Finally, in the third stage we have re-formulated the full
coupled system of equations for the liquid film height and surfactant
concentration profile in a gradient dynamics form, based on an underlying free
energy functional that accounts for wettability, capillarity and
entropic contributions for the surfactant. The resulting equations are
equivalent to the hydrodynamic form for the case of a linear equation
of state for the surfactant.

Based on this thermodynamic re-formulation, we have proposed
amendments to the basic hydrodynamic model that account for four
different physical effects that all may be included through changes to
the free energy functional.  In particular, we have extended the thin-film model to
consistently account for (i) nonlinear equations of state, (ii)
surfactant-dependent wettability, (iii) surfactant phase transitions,
and (iv) substrate-mediated condensation.  The ideas that we have presented can also
be directly applied to films covered by monolayers of nano-sized
particles that are not soluble in the liquid film \cite{Fain06}, or any substance
that remains on the surface of the liquid film.

Our results indicate that nearly all long-wave models found in the
literature that extend the hydrodynamic equations for thin liquid films covered
by insoluble surfactants by including non-linear equations of state are either
not fully consistent or not complete.
The most important differences between our model and those in
the literature, as discussed
above in section~\ref{sec:filmsurf-ext}, are:
\begin{itemize}
\item[(i)] When incorporating a nonlinear equation of state, most authors
fail to note that one must also amend the surfactant diffusion term
in the governing dynamical equations.
\item[(ii)] To account for a surfactant-dependent wettability, it is
  not sufficient to just adapt the Derjaguin (or disjoining) pressure. The
  Marangoni and diffusion term must also be amended.
\item[(iii)] To account for surfactant phase transitions, square
  gradient, or other non-local terms for the surfactant concentration
  must be incorporated into the free energy functional. In these, the
  gradient should be taken along the free surface.
\item[(iv)] When incorporating terms to describe a surfactant phase transition that
  depend on the distance between the film surface
  to the solid substrate (e.g., to describe substrate
  mediated condensation), the added coupling terms lead to
  additional terms in the equation of state
  \textit{and} the Derjaguin pressure.
\end{itemize}
The corrections to models in the literature that result from following
our approach will in many cases only result in quantitative
(rather than qualitative) changes to the results,
that may also be rather small. In some cases, however, the
differences will be qualitative and significant.  For instance, we believe that the
oscillatory dewetting modes (``dewetting waves'') described in
Ref.~\cite{FiGo07} for one and two-layer films with surfactant, are
present in the model as a
consequence of a broken variational structure of the governing
equations that stems from omitting terms in the equation for the
time evolution of the surfactant concentration profile. Using the complete
equations,
all eigenvalues of the linearised problem are real, and thus all
instability modes are monotonic.

The various extensions that we have proposed may all be simultaneously
included so as to account
for more complex situations. The corresponding free energy functional is
\begin{equation}
F[h,\Gamma]\,
=\,\int\left\{f(h,\Gamma)+g(\Gamma,h)\,\xi
+\frac{\kappa}{2} \left(\nabla\Gamma\right)^2
\frac{1}{\xi} \right\}\,dA
\mylab{eq:en:full}
\end{equation}
where $f(h,\Gamma)$ is a generalised wetting interaction term and
$g(\Gamma,h)$ is a generalised local free energy of the surfactant on the free
surface. Note that future work should identify the
connections that must exist between these more general functions $f$ and $g$
because they both arise from the same molecular interactions;
this is an issue that we have not touched upon here.

Our approach may also be further extended to accommodate more general
terms that one should expect to be present in the free energy functional,
including non-local integral (convolution)
contributions to $F$, which are commonly used in dynamical density
functional theory (DDFT) \cite{MaTa99,MaTa00,ArEv04,ArRa04,Poto11},
which uses as input the free energy functionals coming
from equilibrium density functional theory \cite{Evan79, Evan92}.

We emphasise that points (i) and (iii) above are particularly important for
a number of bio-physical systems such as, e.g., the description of the
surfactant layers that reside on the aqueous thin-film of the lung
lining \cite{JeGr92}, where (a)  the equations of state that are used are
strongly nonlinear and (b) experiments show that phase transitions
frequently occur, e.g., in layers of porcine lung surfactant at the air-water interface
at physiologically relevant concentrations and temperatures
\cite{Nag98}. Similar results are found for calf lung surfactant, where
an expanded-to-condensed phase transition is observed as the surfactant
concentration is increased \cite{Disc96}. The dynamics of the ongoing
surfactant phase transitions and their interaction with the hydrodynamics of the thin
liquid film is highly important. The formulation and extensions we
present here allows one to extend the `classical' hydrodynamic
thin-film models to include the more intricate thermodynamic
effects based on equations of state (obtained from suitable free energy
functionals) that are observed experimentally. Note, however, that a
real layer of lung lining is much more complicated  than the
idealised situations mentioned above \cite{PeWe10}, as it
consists of mixtures of soluble surfactants. Work is currently underway
to extend our approach to describe soluble surfactants \cite{TAP12b}.
The extension towards mixtures of
surfactants is more straightforward, but including a third field in the system
makes the algebra somewhat tedious.

Further extensions that should be considered in the future concern the
dynamical aspects. Here, we have employed the mobility matrix $\vec{Q}$
in Eq.~\eqref{eq:mob}, which is
derived from the well known hydrodynamic transport equations obtained for the
simplest case of an insoluble surfactant with a linear equation of
state. Although we believe this approximation should hold over a large parameter range,
at very high surfactant concentrations one must make
corrections. Although, slip at the solid substrate can easily be
accounted for, it is not clear what changes to the mobility matrix
$\vec{Q}$ should arise from incorporating surface viscosity effects and/or
a no-slip condition at the surfactant-covered surface.

Here we have only discussed the gradient dynamics formulation in the
context of surfactant-covered liquid films on solid substrates, because the
mathematical formulation is most convenient. However, it is important
to note that most of the effects that we mention also occur in other geometries.  A
prominent example where our considerations also apply are soap films
based on insoluble surfactants.  This is important for many
systems, such as those reviewed in \cite{BhRu97}, that involve, for instance,
surfactant-dependent Derjaguin pressures and highly nonlinear
equations of state. Another example are surfactant-covered drops of
liquid immersed in another
fluid where issue (i) is particularly relevant when, for example, studying the
shear-driven deformation and/or breakup of such droplets
\cite{PaSt96,EPS99,KvM04,Hame08} or liquid bridges or threads
\cite{AmBa99,TiLi02,CMP02,LFB06}.
In this case, for instance, incorporating a nonlinear equation of state
should also be accompanied by the corresponding amendment of the
surfactant diffusion term.

\appendix

\section{Variational calculus in the general case}
\mylab{sec:app-vari}

The free energy for the surfactant covered thin liquid film is
\begin{equation}
F[h,\widetilde\Gamma/\xi]\,=\,\int\left\{f(h,\widetilde\Gamma/\xi)+g(\widetilde\Gamma/\xi,h)\,\xi
+\frac{\kappa}{2} \left(\nabla(\widetilde\Gamma/\xi)\right)^2
\frac{1}{\xi} \right\}\,dA.
\mylab{eq:en:full}
\end{equation}
We define
\begin{equation}
F[h,\widetilde\Gamma/\xi] = F_\mathrm{wet} + F_\mathrm{surf} + F_\mathrm{grad}
\end{equation}
so as to be able to calculate the variations of the three terms in the free energy separately.
In the following, we limit ourselves to the 1-dimensional case and we often need 
to use the result:
\begin{equation}
\partial_x \xi = \partial_x \sqrt{1+(\partial_x h)^2} = \frac{1}{\xi}(\partial_x h)(\partial_{xx} h).
\end{equation}

\subsection{Variations with respect to $h$}
\begin{equation}
\frac{\delta F_\mathrm{wet}}{\delta h} = \partial_h f
\end{equation}

\begin{eqnarray}
\frac{\delta F_\mathrm{surf}}{\delta h} &=& \partial_h g - \frac{d}{dx}\left[
  -\xi (\partial_\Gamma g)\widetilde\Gamma\frac{1}{\xi^3}\partial_x h + g \frac{1}{\xi}\partial_x h
\right]\\
&=& \partial_h g - \frac{d}{dx}\left[\frac{1}{\xi}(g
-\Gamma\partial_\Gamma g)\partial_x h\right].
\end{eqnarray}
For the next one we need to use
\begin{equation}
\frac{\delta (\int\star dx)}{\delta h}= \frac{\partial \star}{\partial
  h}-\frac{d}{dx}\frac{\partial \star}{\partial (\partial_x h)}
+\frac{d^2}{dx^2}\frac{\partial \star}{\partial (\partial_{xx} h)}.
\end{equation}
We also need
\begin{equation}
\frac{\partial}{\partial h} \xi =0,
\end{equation}
\begin{equation}
\frac{\partial}{\partial (\partial_x h)}  \xi  =
\frac{1}{\xi}\partial_x h
\qquad\mbox{and}\qquad
\frac{\partial}{\partial (\partial_x h)}  \frac{1}{\xi}  =
-\frac{1}{\xi^3}\partial_x h
\end{equation}
and
\begin{eqnarray}
\partial_x
\frac{\widetilde\Gamma}{\xi}&=&\frac{\partial_x\widetilde\Gamma}{\xi}-\frac{\widetilde\Gamma}{\xi^2}\partial_x\xi\\
&=&\frac{\partial_x\widetilde\Gamma}{\xi}-\frac{\widetilde\Gamma}{\xi^3}(\partial_x h)\partial_{xx} h
\end{eqnarray}

\begin{eqnarray}
  \frac{\delta F_\mathrm{grad}}{\delta h} &=& -\frac{d}{dx}\left[ 
-\frac{\kappa}{2}\left(\partial_x\frac{\widetilde\Gamma}{\xi}\right)^2\frac{\partial_x h}{\xi^3}
-\frac{\kappa}{\xi^4}\left(\partial_x\widetilde\Gamma\partial_x h
+\widetilde\Gamma\partial_{xx} h
-3\frac{\widetilde\Gamma}{\xi^2}(\partial_x h)^2\partial_{xx} h\right)\partial_x\frac{\widetilde\Gamma}{\xi}
\right]\nonumber\\
&&-\frac{d^2}{dx^2}\left[\frac{\kappa}{\xi^3} \left(\partial_x\frac{\widetilde\Gamma}{\xi}\right)\Gamma\partial_x h
\right]\nonumber\\
 &=& -\frac{d}{dx}\left\{ \frac{\kappa}{\xi^3}\left[
-\frac{1}{2}\left(\partial_x \Gamma\right)^2\partial_x h
-\left(\partial_x\Gamma\partial_x h +\Gamma\partial_{xx} h
-2\frac{\Gamma}{\xi^2}(\partial_x h)^2\partial_{xx} h\right)\partial_x\Gamma
\right.\right.\nonumber\\
&&\left.\left.-\left(3\frac{\Gamma}{\xi^2}  (\partial_x  h)^2\partial_{xx} h
-  \partial_x\Gamma\partial_x h
-  \Gamma\partial_{xx} h\right)\partial_x\Gamma
+ \Gamma\partial_x h\partial_{xx}\Gamma
\right]\right\}\nonumber\\
 &=& \frac{d}{dx}\left\{ \frac{\kappa}{\xi^3}\left[
\frac{1}{2}\left(\partial_x \Gamma\right)^2\partial_x h
+\frac{\Gamma}{\xi^2}(\partial_x h)^2(\partial_{xx} h)\partial_x\Gamma
- \Gamma\partial_x h\partial_{xx}\Gamma
\right]\right\}
\end{eqnarray}

\subsection{Variations with respect to $\widetilde\Gamma$}
\begin{equation}
\frac{\delta F_\mathrm{wet}}{\delta\widetilde\Gamma} = \frac{1}{\xi}\partial_\Gamma f
\end{equation}

\begin{equation}
\frac{\delta F_\mathrm{surf}}{\delta\widetilde\Gamma} = \partial_\Gamma g
\end{equation}

\begin{eqnarray}
  \frac{\delta F_\mathrm{grad}}{\delta\widetilde\Gamma} &=& -\frac{\kappa}{\xi^4}(\partial_x\Gamma)(\partial_x h)(\partial_{xx} h)
-\frac{d}{dx}\left[    \frac{\kappa}{\xi^2}\partial_x\Gamma
  \right]\\ &=& -\frac{\kappa}{\xi^4}(\partial_x\Gamma)(\partial_x
  h)(\partial_{xx} h) +\kappa\left[ \frac{2}{\xi^4}(\partial_x
    h)(\partial_{xx} h)(\partial_x\Gamma) -\frac{1}{\xi^2}\partial_{xx}\Gamma
\right]\\
&=&\frac{\kappa}{\xi^4}(\partial_x\Gamma)(\partial_x  h)(\partial_{xx}
h) -\frac{\kappa}{\xi^2}\partial_{xx}\Gamma
\end{eqnarray}

\subsection{Collecting the terms}
The
resulting expressions are
\begin{eqnarray}
\frac{\delta  F}{\delta h} \,&=&\, \partial_h f + \partial_h g - \frac{d}{dx}\left[\frac{1}{\xi}\left(g
-\Gamma\partial_\Gamma g -\frac{\kappa}{2\xi^2}\left(\partial_x \Gamma\right)^2
+ \frac{\kappa}{\xi^2}\Gamma\partial_{xx}\Gamma
\right)\partial_x h
-\frac{\kappa}{\xi^5}\Gamma(\partial_x h)^2(\partial_{xx} h)\partial_x\Gamma
\right]
\nonumber\\
\frac{\delta  F}{\delta \widetilde{\Gamma}}
\,&=&\,\frac{1}{\xi}\partial_\Gamma f + \partial_\Gamma g  -\frac{\kappa}{\xi^2}\partial_{xx}\Gamma
+ \frac{\kappa}{\xi^4}(\partial_x\Gamma)(\partial_x  h)\partial_{xx} h
\mylab{eq:app-vari}
\end{eqnarray}
This seems the appropriate stage in the derivation to apply the long-wave
approximation, i.e., to use $(\partial_x
h)^2\equiv\varepsilon\ll1$. Therefore $\xi\approx1+(1/2)\varepsilon^2$
and one obtains 
\begin{eqnarray}
\frac{\delta  F}{\delta h} \,&=&\, \partial_h f + \partial_h g - \frac{d}{dx}\left[\left(g
-\Gamma\partial_\Gamma g -\frac{\kappa}{2}\left(\partial_x \Gamma\right)^2
+ \kappa\Gamma\partial_{xx}\Gamma
\right)\partial_x h
-\kappa\Gamma(\partial_x h)^2(\partial_{xx} h)\partial_x\Gamma
\right]
\nonumber\\
\frac{\delta  F}{\delta \widetilde{\Gamma}}
\,&=&\,\partial_\Gamma f + \partial_\Gamma g  -\kappa\partial_{xx}\Gamma
+ \kappa(\partial_x\Gamma)(\partial_x  h)\partial_{xx} h
\mylab{eq:app-vari2}
\end{eqnarray}
The respective last term is $O(\varepsilon^2)$ smaller than the other
terms with prefactor $\kappa$ and can therefore safely be dropped,
yielding
\begin{eqnarray}
\frac{\delta  F}{\delta h} \,&=&\, \partial_h f + \partial_h g - \frac{d}{dx}\left[\left(g
-\Gamma\partial_\Gamma g -\frac{\kappa}{2}\left(\partial_x \Gamma\right)^2
+ \kappa\Gamma\partial_{xx}\Gamma
\right)\partial_x h
\right]
\nonumber\\
\frac{\delta  F}{\delta \widetilde{\Gamma}}
\,&=&\,\partial_\Gamma f + \partial_\Gamma g  -\kappa\partial_{xx}\Gamma
\mylab{eq:app-vari3}
\end{eqnarray}
Eqs.~(\ref{eq:nonlocal-variations}) in the main text are obtained by
setting $\partial_h g=\partial_\Gamma f=0$, whereas the results in
Eqs.~(\ref{eq:insolsurf-variations}) are obtained by
setting $\partial_h g=\partial_\Gamma f=0$ together with $\kappa=0$.

\acknowledgments

This work was supported by the European Union under grant
PITN-GA-2008-214919 (MULTIFLOW).


%

\end{document}